\documentclass[twoside]{dis07}
\usepackage[latin1]{inputenc}
\usepackage[dvips]{graphicx,epsfig,color}
\usepackage{wrapfig,rotating}
\usepackage{amssymb,amsmath,array}

\begin{document}

\title{Heavy Quark Mass Effects in PQCD and Heavy Flavor Parton
Distributions }
\author{Wu-Ki Tung$^{1,2}$, H.L.~Lai$^{1,2,3}$, J.~Pumplin$^1$, P.~Nadolsky$%
^4$, and C.-P.~Yuan$^1$. \vspace{.3cm} \\
%EndAName
$^1$ Michigan State University, East Lansing, MI - USA \vspace{.1cm}\\
$^2$ University of Washington, Seattle, Washington - USA \vspace{.1cm}\\
$^3$ Taipei Municipal University of Education, Taipei, Taiwan \vspace{.1cm}\\
$^4$ Argonne National Laboratory, Argonne, IL, USA\\
}
\maketitle

\begin{abstract}
The systematic treatment of heavy quark mass effects in DIS in current CTEQ
global analysis is summarized. Applications of this treatment to the
comparison between theory and experimental data on DIS charm production are
described. The possibility of intrinsic charm in the nucleon is studied. The
issue of determining the charm mass in global analysis is discussed.
\end{abstract}

\section{Introduction}

Contemporary global QCD analyses of high precision Deep Inelastic Scattering
(DIS) data, along with other hard processes, require a consistent treatment
of heavy quark mass effects in the perturbative QCD (PQCD) framework. This
review \cite{TungHQ} summarizes key features of the formalism implemented in
the current CTEQ global analysis project \cite{Tung:2006tb} and results on
its application to heavy flavor physics in global analysis \cite{TungSf}.
Sec.\ \ref{sec:hfp} presents the results of the new global fits compared to
heavy flavor production data in DIS \cite{Tung:2006tb}. Sec.~\ref{sec:IC}
addresses issues related to possible intrinsic charm in the nucleon \cite%
{Pumplin:2007wg}. Sec.~\ref{sec:IC} discusses the topical question: can the
charm mass be reliably determined in global QCD analysis?

Due to space limitation, it is impossible to include in this short written
report the figures that illustrate the results discussed in the corresponding
talk, as summarized above. However, since the slides for the talk have been
made available at the official conference URL \cite{TungHQ}, we shall make use
of these, and refer the reader to the actual figures by the slide numbers where
they appear in the posted talk \cite{TungHQ}. The same space limitation
restricts citations to only the papers and talks on which this report is
directly based.

\section{General PQCD framework including heavy quark masses\label%
{sec:theory}}

The key features of the general-mass PQCD framework of \cite{Tung:2006tb} is
illustrated in slide 3 of \cite{TungHQ}.

\noindent \textbf{Factorization Formula and (scheme-dependent) summation over
parton flavors: } Collins has established that the PQCD factorization theorem
for the structure functions takes the general form $F_{\lambda
}(x,Q^{2})=\sum_{a}f^{a}\otimes \widehat{\omega }_{a}^{\lambda }$ even when the
heavy quark mass effects are kept. Here, the summation is over the active
parton flavor label $a$, $f^{a}(x,\mu )$ are the parton distributions at the
factorization scale $\mu $, and $\widehat{\omega }_{a}^{\lambda }(x,Q/\mu
,M_{i}/\mu )$ are the infrared safe Wilson coefficients (or hard-scattering
amplitudes) that can be calculated order-by-order in perturbation theory. The
summation over \textquotedblleft parton flavor\textquotedblright\ label $a$ in
the factorization formula is determined by the \emph{factorization scheme}
chosen to define the parton distributions $f^{a}(x,\mu )$. In general, we use
the variable flavor number scheme.

\noindent \textbf{The summation over (physical) final-state flavors:} For
total inclusive structure functions, the factorization formula contains an
implicit summation over all possible quark flavors in the final state: $\hat{%
\omega}_{a}=\sum_{b}\hat{\omega}_{a}^{b}$, where \textquotedblleft $b$%
\textquotedblright\ denotes final state flavors, and $\hat{\omega}_{a}^{b}$
is the perturbatively calculable hard cross section for an incoming parton
\textquotedblleft $a$\textquotedblright\ to produce a final state containing
flavor \textquotedblleft $b$ \textquotedblright . It is important to
emphasize that \textquotedblleft $b$\textquotedblright\ labels quark flavors
that can be produced \emph{physically} in the final state; it is \emph{not}
a parton label in the sense of initial-state parton flavors described in the
previous subsection. In a proper implementation of the general-mass (GM)
formalism, the distinction between the initial-state and final-state
summations must be unambiguously and correctly observed.

\noindent \textbf{Kinematic constraints and rescaling:} Kinematic
constraints from the phase space treatment have a significant impact on the
numerical results of the calculation. In DIS, with heavy flavor produced in
the final state, the most natural way to ensure the correct kinematics for
both NC and CC processes is to use the \emph{rescaling variable }$\chi =x
(1+( \Sigma _{f}~M_{f}/Q)^{2}) $ in place of the usual Bjorken $x$ in the
convolution integral of the factorization formula. Here $\Sigma _{f}~M_{f}$
is the sum of all heavy flavor masses in the final state. This is the ACOT$%
\chi $ prescription used in most recent literature.

\noindent \textbf{Hard Scattering Amplitudes and the SACOT Scheme:} The hard
scattering amplitude $\widehat{\omega }_{a}^{\lambda }\left( x,Q/\mu
,M_{i}/\mu \right) $ is by definition infrared safe, meaning it is free
from logarithmic \textquotedblleft mass-singularities" in the limit $%
M_{i}/Q\rightarrow 0$. Within the PQCD formalism, there is some freedom to
choose how the finite mass effects are treated. The choice that makes the
calculation simplest while retaining full accuracy (the SACOT scheme) can be
stated as: keep the heavy quark mass dependence in the Wilson coefficients for
partonic subprocesses with only light initial state partons ($g,u,d,s$); but
use the zero-mass Wilson coefficients for subprocesses that have an initial
state heavy quark ($c,b$). For the 4-flavor scheme to order $\alpha _{s}$
(NLO), we do the following: (a) keep the full $M_{c}$ dependence of the gluon
fusion subprocess; (b) for NC scattering ($\gamma /Z$ exchanges), set all quark
masses to zero in the quark-initiated subprocesses; and (c) for CC scattering
($W^{\pm }$ exchange), set the initial-state quark masses to zero, but keep the
final-state quark masses on shell.

\noindent \textbf{Choice of Factorization Scale:} The total inclusive
structure function $F_{i}^{tot}$ is infrared safe. Consider the simple case
of just one effective heavy flavor charm, $%
F_{i}^{tot}=F_{i}^{light}+F_{i}^{c}$ for any given flavor-number scheme. Since
the right-hand side of this equation is dominated by the light-flavor term
$F_{i}^{light}$, and the natural choice of scale for this term is $\mu =Q$, it
is reasonable to use this choice for both terms to ensure infrared safety. On
the other hand, in the case of experimentally measured semi-inclusive DIS
structure functions for producing a charm particle in the final state,
$F_{i}^{c}$ is theoretically \emph{infrared unsafe} beyond NLO. One may
nonetheless perform comparison of NLO theory with experiment with the
understanding that the results are intrinsically less reliable, and they can be
sensitive to the choice of parameters. The
most natural choice of factorization scale in this case is $\mu =\sqrt{%
Q^{2}+M_{c}^{2}}$.

\section{Results and Comparison with heavy flavor production data \label%
{sec:hfp}}

Slides 4 and 5 of \cite{TungHQ} show the size of heavy quark mass effects on
the calculation of $F_{2}(x,Q)$ and $F_{L}(x,Q)$. The color coded areas (with
complementary contours) indicate the fractional differences between GM and
zero-mass (ZM) calculations. Understandably, the largest differences occur at
low $Q$ and low $x$; and the significance is much more for $F_{L}(x,Q)$ than
for $F_{2}(x,Q)$, since the former vanishes at LO for the ZM case. As indicated
in slide 6, the GM calculation is stable and robust. It has been used as the
basis for a new round of global analysis of PDFs, using the full set of HERA
Run I neutral current (NC) and charged current (CC) total cross section and
heavy flavor production data, along with the usual DY and jet data
cf.~\cite{TungSf}. Here we shall only present the comparison of the new fits to
the heavy flavor production data measured at HERA.

Slide 10 shows the comparison of the ZEUS 1996-97 and 1998-2000 charm
production data to the theory values obtained with the new PDF sets CTEQ6.5M
(same shape for strange and non-strange seas, \cite{Tung:2006tb}), CTEQ6.5S0
(independent shapes for strange and non-strange seas, \cite{Lai:2007dq}) as
well as for the older CTEQ6HQ. Plotted are ratios of $F_{2}^{c}(x,Q)$ to
that of a best fit to the respective data set. \ The fits to data are all
reasonable. The new PDFs give slightly better fits than the previous one. \
Slide 11 shows the comparison of the H1 charm and bottom production data to
the theory values from the same PDF sets. The $F_{2}^{c}(x,Q)$ data points
have more scatter around the (smooth) theory values. The overall $\chi ^{2}$
of these fits is however acceptable. \

It is worth noting that correlated systematical errors are always taken into
account in our global analysis. The data points shown on these plots have
been shifted by the fitted systematic errors; hence the differences between
the data points and the theory values as they appear on these plots give a
faithful indication of the quality of the fits.

\section{Is there intrinsic charm in the nucleon?\label{sec:IC}}

Many nonperturbative models of nucleon structure suggest the existence of
\emph{intrinsic charm} (IC)---a non-vanishing component of nucleon parton
structure at the scale of $M_c$. On the other hand, practically all global
analysis of the parton structure of the nucleon so far ignore this possibility
and make the simplifying assumption that all heavy quark partons are
radiatively generated: they only arise from perturbative QCD evolution,
starting from zero at $\mu \sim M_c$. Where does the truth lie? The resolution
of this dichotomy is of inherent physics interest because it concerns the
fundamental structure of matter, as well as of practical interest because the
cross sections for many beyond-the-standard-model (BSM) processes at hadron
colliders depend on the charm parton content of the nucleon. We have addressed
this problem phenomenologically by a careful global analysis based on the GM
PQCD formalism that, for the first time, allows for an independent charm sector
\cite{Pumplin:2007wg}.

As indicated in slide 14 of \cite{TungHQ}, the following specific scenarios for
the charm sea, $c(x,\mu =M_{c})$, are explored within our GM global analysis
framework: (i) the conventional radiatively generated charm; (ii) non-vanishing
IC $c(x,M_{c})$ that is \emph{sea-like }(i.e. shaped as the light
sea quarks); and (iii) IC of the kind suggested by light-cone wave
function models of the nucleon (peaked at moderately large $x$). Within
scenario (iii), we further distinguish two models: the one studied by Brodsky
et al.~(the BHPS model), and a meson cloud model.

Slide 15 summarizes the main results. The figure shows the goodness-of-fit
for the global analysis, $\chi _{global}^{2}$, as a function of the magnitude
of the IC component, measured by the momentum fraction carried $%
\langle x\rangle _{c+\bar{c}}$, under the various scenarios. In the range $%
0<\langle x\rangle _{c+\bar{c}}<0.01$ (outlined by the horizontal oval), $%
\chi _{global}^{2}$ is largely insensitive to $\langle x\rangle _{c+\bar{c}}$%
, indicating that there is no strong evidence for or against IC
of a magnitude in this range. \ However, outside this range, for $\langle
x\rangle _{c+\bar{c}}>0.01$ (outlined by the vertical oval), we see a
precipitous rise of $\chi _{global}^{2}$ as $\langle x\rangle _{c+\bar{c}}$
increases. Thus our global analysis sets a useful upper bound on the amount
of intrinsic charm that is consistent with existing data. Using a 90\%
confidence level (C.L.) criterion, this bound is $\langle x\rangle _{c+\bar{c%
}}<0.02$.

Although models of IC generally do not predict $\langle x\rangle _{c+\bar{c}}$,
typical guesstimates place it around $0.01$. This is consistent with the bound
we determined from the above global analysis. The presence of IC of such a
magnitude can have an impact on certain BSM processes, such as charged Higgs
production in hadron collider phenomenology. Cf.~slide 7 of \cite{TungSf}, and
\cite{YuanEw,Lai:2007dq}.
Slide 16 shows the charm distribution $c(x,\mu )$ at three energy scales $%
\mu =1.3,~3.16,~85$ GeV in the BHPS scenario, for various magnitudes of the
initial distribution. We see the radiatively generated component (peaked at
small $x$) catching up with the IC component (peaked at moderate $x$)
as $\mu $ increases. However, the latter clearly still dominates in the $x$
region $\geqq 0.1$ even at the W/Z mass scale.

\section{Can the charm mass be determined in global analysis?\label{sec:ChM}}

In principle, heavy quark masses $M_{i}(\mu )$ at some renormalization scale
$\mu $ are basic parameters of QCD, similar to the coupling $\alpha _{s}(\mu
)$. Thus, just like for $\alpha _{s}$, there has been recent interest in
determining $M_{i}$, e.g.~the charm mass $M_{c}$, from global QCD analysis. In
particular, is it possible to perform a conventional global QCD analysis using
$M_{c}$ as one of the fitting parameters, and thereby determine the charm mass
to be the one that gives the best fit? If so, one may further ask, is this mass
the $\overline{MS}$-mass or the pole-mass?

Slides 18-20 show results of a study, following the above procedure literally:
one finds that the global analysis favors a relatively small values of
$M_{c}\sim 1.3$ GeV, and the goodness-of-fit $\chi _{global}^{2}$ increases
with $M_{c}$. But, a closer examination of the problem immediately raises the
question: what is the physical meaning of this favored value of $M_{c}$? The
problem is, a chosen value of $M_{c}$ affects the global QCD analysis in two
distinct ways: (i) through the mass-dependent Wilson coefficients in the
theoretical calculation (the pole-mass); and (ii) through the initial condition
$c(x,\mu =M_{c})=0$---the implicit assumption of \emph{radiatively generated
charm} that is used in all existing global analyses. It turns out, the global
fit is influenced much more by the latter than by the former. Since radiatively
generated charm is only an assumption, not an integral part of the QCD theory,
the value of $M_{c}$ favored by global analysis is not
directly related to the basic QCD charm mass parameter---it is neither the $%
\overline{MS}$-mass nor the pole-mass! In order to answer the original question
\textquotedblleft can the charm mass be determined in global
analysis?\textquotedblright, one needs to clearly differentiate between the
two sources of dependence on $M_{c}$ mentioned above. This is currently under study.

\medskip %\section{Bibliography}

% ****************************************************************************
% BIBLIOGRAPHY AREA
% ****************************************************************************

% ****************************************************************************
% END OF BIBLIOGRAPHY AREA
% ****************************************************************************


\begin{thebibliography}{9}
\bibitem{TungHQ} {\footnotesize This review is based on the talk of Wu-Ki
Tung in the Heavy Flavor workgroup of DIS07:\newline
\verb|http://indico.cern.ch/contributionDisplay.py?contribId=276&sessionId=5&confId=9499|%
.}

\bibitem{TungSf} {\footnotesize Slides: See Wu-Ki Tung, talk in the Structure
    Function workgroup in these proceedings:\newline
\verb|http://indico.cern.ch/contributionDisplay.py?contribId=189&sessionId=8&confId=9499|%
.}

\bibitem{Tung:2006tb} {\footnotesize W.~K.~Tung, H.~L.~Lai, A.~Belyaev,
J.~Pumplin, D.~Stump and C.~P.~Yuan,
%``Heavy quark mass effects in deep inelastic scattering and global QCD
%analysis,''
JHEP \textbf{0702}, 053 (2007).}

\bibitem{YuanEw} {\footnotesize See also C.-P.~Yuan, talk in the Electroweak
    work group in these proceedings:\newline
\verb|http://indico.cern.ch/contributionDisplay.py?contribId=120&sessionId=9&confId=9499|%
.}

\bibitem{Lai:2007dq} {\footnotesize H.~L.~Lai, P.~Nadolsky, J.~Pumplin,
D.~Stump, W.~K.~Tung and C.~P.~Yuan,
%``The Strange Parton Distribution of the Nucleon: Global Analysis and
%Applications,''
JHEP \textbf{0704}, 089 (2007).}

\bibitem{Pumplin:2007wg} {\footnotesize J.~Pumplin, H.~L.~Lai and
W.~K.~Tung, %``The charm parton content of the nucleon,''
Phys.\ Rev.\ D \textbf{75}, 054029 (2007).}
\end{thebibliography}
\end{document}